\documentclass[aps,prd,nofootinbib,preprintnumbers,twocolumn]{revtex4}

\usepackage{bm}
\usepackage{latexsym}
\usepackage{dcolumn}
\usepackage{amsfonts,amssymb}
\usepackage{graphicx,epsfig}
\usepackage{psfrag}
\usepackage{amsmath,amssymb}
\usepackage{srcltx}
\usepackage{color}
\usepackage[normalem]{ulem}
\usepackage{multirow}

\def\beq{\begin{equation}}
\def\eeq{\end{equation}}
\def\br{\begin{eqnarray}}
\def\er{\end{eqnarray}}
\def\benu{\begin{enumerate}}
\def\eenu{\end{enumerate}}
\def\nn{\nonumber} 
\def\f{\frac}
\def\pa{{\partial}}
\def\l{\left}
\def\r{\right}    
\def\d{{\rm d}}

\newcommand{\viz}{\textit{viz.~}}
\newcommand{\ie}{\textit{i.e.~}}

\begin{document}

\title{Moving mirrors and the fluctuation-dissipation theorem}
\author{D.~Jaffino Stargen}
\email[]{E-mail: jaffino@physics.iitm.ac.in}
\author{Dawood~Kothawala}
\author{L.~Sriramkumar}
\affiliation{Department of Physics, Indian Institute of Technology 
Madras, Chennai~600036, India.}
\begin{abstract}
We investigate the random motion of a mirror in $(1+1)$-dimensions that is 
immersed in a thermal bath of massless scalar particles which are interacting 
with the mirror through a boundary condition.
Imposing the Dirichlet or the Neumann boundary conditions on the moving mirror, 
we evaluate the mean radiation reaction force on the mirror and the correlation 
function describing the fluctuations in the force about the mean value.
From the correlation function thus obtained, we explicitly establish the 
fluctuation-dissipation theorem governing the moving mirror.
Using the fluctuation-dissipation theorem, we compute the mean-squared 
displacement of the mirror at finite and zero temperature.
We clarify a few points concerning the various limiting behavior of the 
mean-squared displacement of the mirror. 
While we recover the standard result at finite temperature, we find that the 
mirror diffuses logarithmically at zero temperature, confirming similar
conclusions that have been arrived at earlier in this context. 
We also comment on a subtlety concerning the comparison between zero temperature 
limit of the finite temperature result and the exact zero temperature result.
\end{abstract}
\maketitle


\section{Introduction}

Brownian motion refers to the random motion of small particles immersed in 
a large bath. 
Classic examples of Brownian motion would include the random motion of 
particles floating in a liquid and the motion of dust particles illuminated 
by a ray of sunlight. 
The motion of a Brownian particle is effectively described by the Langevin 
equation (see, for instance, Ref.~\cite{reif-1965}). 
In the Langevin equation, the force experienced by the particle is decomposed 
into two components: one, an averaged force which is dissipative in nature, 
and another that is rapidly fluctuating.
The combination of the dissipative and the fluctuating forces leads to the 
diffusion of the Brownian particle through the bath. 

\par

The amplitude of the dissipative force and the correlation function describing
the fluctuating component are related by the fluctuation-dissipation theorem 
(see, for example, Refs.~\cite{callen-1951,kubo-1957,kubo-1966}).
The theorem can also be utilized to evaluate the mean-squared displacement of 
the Brownian particle and thereby illustrate the diffusive nature of the 
particle. 
It is well known that, in a bath maintained at a finite temperature, the 
mean-squared displacement of the Brownian particle grows linearly with  
time at late times.
An interesting question that seems worth addressing is whether the Brownian 
particle diffuses even at zero temperature (in this context, see
Refs.~\cite{sinha-1992,jaekel-1993a,jaekel-1993b}).

\par

A point mirror moving in a thermal bath provides a splendid example for studying 
these issues and, in fact, the system has been considered earlier in different 
contexts (see, for instance, Refs.~\cite{jaekel-1992,gour-1999,alves-2003,wu-2005,
alves-2008,wang-2014,wang-2015}; also see the following 
reviews~\cite{jaekel-1997,dodonov-2001}).
Our goal in this work is to reconsider the random motion of the mirror immersed
in a thermal bath.
Specifically, our aims can be said to be two-fold.
Our first goal is to evaluate the average force on the moving mirror as well as 
the correlation function characterizing the fluctuating component and explicitly 
establish the fluctuation-dissipation theorem relating these quantities. 
Our second aim is to utilize the fluctuation-dissipation theorem to calculate  
the mean-squared displacement of the mirror both at finite and zero temperature
and, in particular, examine the nature of diffusion at zero temperature.
In order for the problem to be analytically tractable, as is usually done
in this context, we shall work in $(1+1)$-spacetime dimensions and assume
that the mirror is interacting with a massless scalar field (for the original 
discussion, see Refs.~\cite{moore-1970,dewitt-1975,fulling-1976,ford-1982}).
Importantly, one finds that, under these simplifying assumptions, it proves 
to be possible to calculate all the quantities involved explicitly using the 
standard methods of quantum field theory.

\par

A few clarifying remarks concerning the prior efforts in these directions are 
in order at this stage of our discussion.
The earliest efforts in the literature had primarily focussed on carrying out 
the quantum field theory of a massless scalar field in the presence of a moving 
mirror in $(1+1)$-spacetime dimensions~\cite{moore-1970,dewitt-1975}. 
It was immediately followed by efforts to evaluate the regularized stress-energy 
tensor associated with the quantum field in the vacuum state, \ie\/ at zero 
temperature~\cite{fulling-1976,ford-1982}.
These efforts had also arrived at the corresponding radiation reaction force on 
the moving mirror. 
About a decade after these initial efforts, it was recognized that the system 
provides a tractable scenario to examine the validity of the 
fluctuation-dissipation theorem and the behavior of the mean-squared displacement 
of the mirror.
The fluctuation-dissipation theorem at zero temperature was established in this
context and the behavior of the mean-squared displacement of the mirror at large 
times was also arrived at~\cite{jaekel-1992,jaekel-1997}.
More recently, the radiation reaction force on the moving mirror at a finite 
temperature has been calculated as well (in this context, see Ref.~\cite{alves-2008}). 
However, to the best of our knowledge, this is the first time that the 
correlation function governing the radiation reaction force is being
evaluated and the associated fluctuation-dissipation theorem is being 
explicitly established for the case of the moving mirror at a finite 
temperature (though we should clarify that the possibility has been 
briefly discussed in Ref.~\cite{jaekel-1993b}).
Moreover, we believe this is the first effort towards obtaining complete
analytical expressions for the mean-squared displacements of the mirror 
that is valid at all times.

\par

This article is organized as follows. 
In the following section, we shall quickly review the quantization of a 
massless scalar field in the presence of a moving mirror in $(1+1)$-spacetime
dimensions, and evaluate the regularized stress-energy tensor of the scalar 
field at a finite temperature.
We shall use the result to arrive at the radiation reaction force on
the moving mirror. 
In Sec.~\ref{sec:fc}, we shall evaluate the correlation function governing
the fluctuating component of the radiation reaction force.
Using the radiation reaction force and the correlation function characterizing 
the fluctuating component, we shall establish the fluctuation-dissipation 
theorem in Sec.~\ref{sec:fdt}.
In Secs.~\ref{sec:msd-ft} and~\ref{sec:msd-zt}, using the fluctuation-dissipation 
theorem, we shall calculate the mean-squared displacement of the mirror at finite 
and zero temperature. 
Finally, in Sec.~\ref{sec:d}, we shall close with a brief discussion on the 
results we have obtained.
We shall relegate the details concerning some of the calculations to four
appendices.
Specifically, in the final appendix, we shall clarify a subtle point concerning the 
zero temperature limit of the finite temperature result for the mean-squared 
displacement of the mirror.

\par

Note that we shall work in units such that $c=\hbar=k_{_{\rm B}}=1$.
An overdot shall denote differentiation with respect to the Minkowski
time coordinate.
Unless we mention otherwise, overprimes above functions shall represent 
differentiation of the functions with respect to their arguments.
Angular brackets shall, in general, denote expectation values evaluated 
at a finite temperature, barring in Sec.~\ref{sec:msd-zt}, where it 
shall represent expectation values at zero temperature (\ie in the 
quantum vacuum).
Lastly, subscripts and superscripts ${\rm R}$ and ${\rm L}$ shall denote 
quantities to the right and the mirror, respectively. 


\section{Radiation reaction on a mirror moving in a thermal 
bath}\label{sec:qscfmm}

In this section, we shall first discuss the quantization of a massless
scalar field in $(1+1)$-spacetime dimensions in the presence of a moving 
mirror.
We shall impose Dirichlet or Neumann boundary conditions on the mirror
and evaluate the regularized stress-energy tensor for the scalar field
at a finite temperature.
From this result, we shall obtain the radiation reaction force on the 
mirror.


\subsection{Boundary conditions, modes and quantization}

Consider a massless scalar field, say, $\phi$, which is governed by the 
following equation in $(1+1)$-dimensional flat spacetime:
\beq
\f{\pa^2 \phi}{\pa t^2}-\f{\pa^2\phi}{\pa x^2}=0.
\eeq
Let a mirror be moving along the trajectory $x=z(t)$, such that 
$\vert{\dot z}(t)\vert<1$, and let us assume that the scalar field $\phi$ 
satisfies either the Dirichlet or the Neumann boundary conditions on the 
moving mirror.
In the case of the Dirichlet boundary condition, we require that
\beq
\phi\l[t,x=z(t)\r]=0,\label{eq:dbcmm}
\eeq
whereas, in the case of the (covariant) Neumann condition, we shall require
\beq
n^i\, \nabla_i \phi\,\biggr\vert_{x=z(t)}=
\l(\f{\pa \phi}{\pa x}+{\dot x}\,\f{\pa \phi}{\pa t}\r)_{x=z(t)}=0,
\label{eq:nbcmm}
\eeq
where $n^i$ is the vector normal to the mirror trajectory~$z(t)$. 
The mirror divides the spacetime into two completely {\it independent}\/  
regions, to the left (L) and the right (R) of the mirror. 

\par

Let $u_{\omega}^{{\rm R}}(t, x)$ and $u_{\omega}^{{\rm L}}(t, x)$
denote the normalized modes of the scalar field in the regions to 
the right and the left of the mirror, respectively.
\begin{figure}
\includegraphics[width=8.5cm]{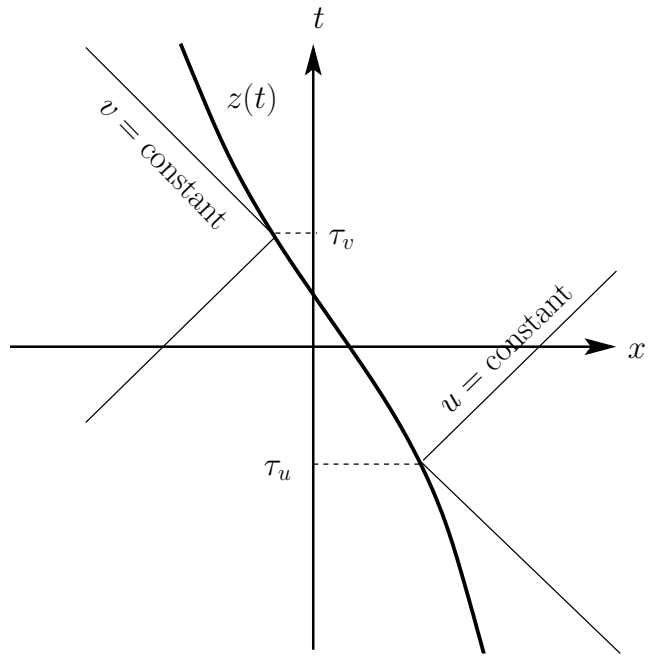}
\caption{The mirror moving along the trajectory $z(t)$ divides the 
spacetime into two distinct regions to the right and the left of 
the mirror.
Note that $\tau_u$ and $\tau_v$ denote the times when the incoming
waves are reflected by the mirror and converted into outgoing waves 
to the right and the left of the mirror, respectively.}
\end{figure}
These modes can be expressed in terms of the null coordinates $u
=t-x$ and $v=t+x$ as follows~\cite{moore-1970,dewitt-1975,
fulling-1976,alves-2003,alves-2008}:
\begin{subequations}
\label{eq:m}
\begin{eqnarray}
u_{\omega}^{_{\rm R}}(t, x)
&=&\f{1}{\sqrt{4\,\pi\, \omega}}\,
\l[\kappa\, {\rm e}^{-i\,\omega\, v} 
+\kappa^\ast\, {\rm e}^{-i\,\omega\, p_1(u)}\r],\label{eq:mr}\\
u_{\omega}^{_{\rm L}}(t, x)
&=&\f{1}{\sqrt{4\,\pi\, \omega}}\,
\l[\kappa\, {\rm e}^{-i\,\omega\, u}
+\kappa^\ast\, {\rm e}^{-i\,\omega\,p_2(v)}\r].\label{eq:ml}
\end{eqnarray}
\end{subequations}
The functions $p_1(u)$ and $p_2(v)$ are given by
\begin{subequations}
\begin{eqnarray}
p_1(u)&=& 2\,\tau_u-u,\\
p_2(v)&=& 2\,\tau_v-v,
\end{eqnarray}
\end{subequations}
where $\tau_{u}$ and $\tau_{v}$ denote the times at which the null lines 
$u$ and $v$ intersect the mirror's trajectory to the right and the left of
the mirror (see accompanying figure).
The quantities $\tau_{u}$ and $\tau_{v}$ are determined by the conditions 
$\tau_u-z(\tau_u)=u$ and $\tau_v+z(\tau_v)=v$.
The quantity $\kappa$ is a constant and its value depends on the boundary
condition, with $\kappa=i$ and $\kappa=1$ corresponding to the Dirichlet 
and the Neumann conditions.

\par

On quantization, the scalar field operator ${\hat \phi}$ to the right
and the left of the mirror can be decomposed in terms of the corresponding
normal modes as follows:
\begin{equation}
{\hat \phi}(t, x)
=\int\limits_{0}^{\infty}\d\omega\,
\l[{\hat a}_{\omega}\, u_{\omega}(t, x)
+{\hat a}_{\omega}^{\dag}\, u_{\omega}^\ast(t, x)\r],\label{eq:phi-d}
\end{equation}
where ${\hat a}_{\omega}$ and ${\hat a}_{\omega}^{\dag}$ are the annihilation 
and the creation operators which obey the standard commutation relations.
It should be emphasized that there exist a separate set of operators defining
the vacuum state and characterizing the corresponding Fock space on either side 
of the mirror.
  

\subsection{Stress-energy tensor at finite temperature}\label{sec:rrftb}

Let us now turn to the evaluation of the expectation value of the stress-energy 
tensor of the quantum scalar field at a finite temperature $T$.
In $(1+1)$-dimensions, the different components of stress-energy tensor are
given by
\begin{subequations}
\label{eq:set}
\begin{eqnarray}
T_{00}&=& T_{11}
=\f{1}{2}\, \l[\l(\f{\pa \phi}{\partial t}\right)^2
+\l(\f{\partial \phi}{\pa x}\r)^2\r],\\
T_{01}&=& T_{10}
=\f{1}{2}\, \l[\f{\pa \phi}{\pa t}\,\f{\pa \phi}{\pa x}     
+\f{\pa \phi}{\pa x}\,\f{\pa \phi}{\pa t}\r],
\end{eqnarray}
\end{subequations}
with the indices $(0,1)$ and corresponding to the spacetime coordinates $(t,x)$.
It is now a matter of substituting the decomposition~(\ref{eq:phi-d}) of the
quantum scalar field in the above expression for the stress-energy tensor and
evaluating the expectation values at a finite temperature $T$ on either side
of the mirror.
All the expectation values can be arrived at from the basic result (see, for
instance, Ref.~\cite{itzykson-1980})
\beq
\langle {\hat a}_{\omega}^{\dag}\, {\hat a}_{\omega'}\rangle
=\f{\delta^{(1)}(\omega-\omega')}{{\rm e}^{\beta\,\omega}-1},
\eeq
where $\beta=1/T$ denotes the inverse temperature. 

\par

Since the stress-energy tensor involves two-point functions in the coincidence
limit, as is well known, one will encounter divergences in calculating the 
quantity (see, for example, Ref.~\cite{birrell-1982}).
In flat spacetime, these divergences correspond to contributions due to the
Minkowski vacuum and they can be easily identified and regularized using, say,
the method of point-splitting regularization~\cite{fulling-1976}.
The regularized stress-energy tensor to the right and the left of the mirror
can be obtained to be
\begin{subequations}
\begin{eqnarray}
\langle\hat{T}^{00}_{_{\rm R}}\rangle 
&=&-\frac{1}{24\,\pi}
\l[\f{p_1'''(u)}{p_1'(u)}
-\f{3}{2}\,\l(\frac{p_1''(u)}{p_1'(u)}\r)^2\right]\nn\\
& &+\,\frac{\pi}{12\,\beta^2}\, \l[1+p_1'^2(u)\r],\label{emr00}\\
\langle\hat{T}^{01}_{_{\rm R}}\rangle 
&=&-\frac{1}{24\,\pi}\,
\l[\f{p_1'''(u)}{p_1'(u)}
-\f{3}{2}\,\l(\f{p_1''(u)}{p_1'(u)}\r)^2\r]\nn\\
& &-\,\f{\pi}{12\,\beta^2}\,\l[1-p_1'^2(u)\r],\label{emr01}\\
\langle\hat{T}^{00}_{_{\rm L}}\rangle 
&=&-\f{1}{24\,\pi}\,
\l[\f{p_2'''(v)}{p_2'(v)}
-\f{3}{2}\, \l(\frac{p_2'''(v)}{p_2'(v)}\r)^2\r]\nn\\
& &+\,\frac{\pi}{12\,\beta^2}\l[1+p_2'^2(v)\r],\label{eml00}\\
\langle\hat{T}^{01}_{_{\rm L}}\rangle 
&=&\f{1}{24\,\pi}\,
\l[\frac{p_2'''(v)}{p_2'(v)}
-\f{3}{2}\, \l(\f{p_2''(v)}{p_2'(v)}\r)^2\right]\nn\\
& &+\,\f{\pi}{12\,\beta^2}\l[1-p_2'^2(v)\r],\label{eml01}
\end{eqnarray}
\end{subequations}
where, recall that, the overprimes denote differentiation of the functions
with respect to the arguments.
Three points concerning the above expressions require emphasis.
To begin with, note that, the terms appearing in the first line of the
above expressions for the components of the stress-energy tensor are  
independent of $\beta$.
These terms represent the vacuum contribution~\cite{fulling-1976}, while
the terms appearing in the second lines are the contributions arising 
due to the finite temperature.
It should be mentioned here that the finite temperature terms include the 
contributions that arise even in the absence of the mirror.
Secondly, note that the stress-energy tensor is a function only of $u$ 
and $v$ to the right and the left of the mirror, respectively.
The moving mirror excites the scalar field and the terms that depend on 
$p_1(u)$ and $p_2(v)$ describe the stress-energy associated with the 
radiation emitted by the mirror due to its motion.
Evidently, the vacuum contribution can be considered as spontaneous emission 
by the mirror, while the finite temperature contributions can be treated as 
stimulated emission. 
Thirdly, one finds that the stress-energy tensor is completely independent
of the boundary condition (actually it depends on $\vert\kappa\vert^2$,
which is unity for the Dirichlet and the Neumann conditions).   

\par

The quantities $p_1(u)$ and $p_2(v)$ and their derivatives with respect to
their arguments can be expressed in terms of the velocity of the mirror
and its two time derivatives.
It can be shown that the components of the stress-energy tensor can be 
expressed in terms of ${\dot z}$, ${\ddot z}$ and ${\dddot z}$ as
follows:
\begin{subequations}
\begin{eqnarray}
\langle\hat{T}^{00}_{_{\rm R}}\rangle
\!\!&=&\!\!-\f{1}{12\,\pi}\l[\frac{\dddot z}{(1-{\dot z})^2\,(1-{\dot z}^2)}
+\f{3\,{\dot z}\,{\ddot z}^2}{(1-{\dot z})^2\,(1-{\dot z}^2)^2}\r]\nn\\
& &+\,\frac{\pi}{6\,\beta^2}\,\f{1+{\dot z}^2}{(1-{\dot z})^2},\label{0emr00}\\
\langle\hat{T}^{01}_{_{\rm R}}\rangle
\!\!&=&\!\!-\f{1}{12\,\pi}\,\l[\f{\dddot z}{(1-{\dot z})^2\,(1-{\dot z}^2)}
+\frac{3\,{\dot z}\,{\ddot z}^2}{(1-{\dot z})^2\,(1-{\dot z}^2)^2}\r]\nn\\
& &+\,\frac{\pi}{3\,\beta^2}\,\f{\dot z}{(1-{\dot z})^2},\label{0emr01}\\
\langle\hat{T}^{00}_{_{\rm L}}\rangle
\!\!&=&\!\!\f{1}{12\,\pi}\,\l[\f{\dddot z}{(1+{\dot z})^2\,(1-{\dot z}^2)}
+ \frac{3\,{\dot z}\,{\ddot z}^2}{(1+{\dot z})^2\,(1-{\dot z}^2)^2}\r]\nn\\
& &+\,\frac{\pi}{6\,\beta^2}\,\f{1+{\dot z}^2}{(1+{\dot z})^2},\label{0eml00}\\
\langle\hat{T}^{01}_{_{\rm L}}\rangle
\!\!&=&\!\!-\f{1}{12\,\pi}\,\l[\f{\dddot z}{(1+{\dot z})^2\,(1-{\dot z}^2)}
+\f{3\,{\dot z}\,{\ddot z}^2}{(1+{\dot z})^2\,(1-{\dot z}^2)^2}\r]\nn\\
& &+\,\f{\pi}{3\,\beta^2}\,\f{\dot z}{(1+{\dot z})^2},\label{0eml01}
\end{eqnarray}
\end{subequations}
where the velocity and its time derivatives are to be evaluated at the 
retarded times (\ie  $\tau_u$ or $\tau_v$) when the radiation was emitted 
by the mirror.
At this stage, it is useful to note that, while the vacuum terms depend
on the velocity ${\dot z}$, the acceleration ${\ddot z}$ as well as the time
derivative of the acceleration ${\dddot z}$~\cite{fulling-1976,ford-1982}, 
the finite temperature term involves only the velocity ${\dot z}$.


\subsection{Radiation reaction force on the moving mirror}

The energy emitted by the moving mirror due to its interaction with the 
scalar field leads to a radiation reaction force on the mirror. 
The radiation reaction force can be obtained from the conservation of the
total momentum of the mirror and the scalar field.
The operator describing the radiation reaction force on the mirror can be 
expressed as~\cite{ford-1982}		 
\begin{eqnarray}
\label{mp}
{\hat F}_{\rm rad}=-\frac{\d {\hat P}^x}{\d t},
\end{eqnarray}
where ${\hat P}^x$ is the momentum operator associated with the scalar 
field and is given by
\begin{equation}
{\hat P}^x\equiv
\int_{-\infty}^{z(t)}\d x\, {\hat T}_{_{\rm L}}^{01}
+\int_{z(t)}^{\infty}\d x\, {\hat T}_{_{\rm R}}^{01}.\label{eq:Px-op}
\end{equation}
The mean value of the radiation reaction force, evaluated at a finite
temperature, can be arrived at from the expectation values of the 
stress-energy tensor we have obtained above.
One finds that, the mean radiation reaction force can be expressed as
\begin{eqnarray}
\langle {\hat F}_{\rm rad}\rangle
&=&\f{1}{6\,\pi}\,\f{1}{(1-{\dot z}^2)^{1/2}}\,
\f{\d}{\d t}\l[\f{\ddot z}{\l(1-{\dot z}^2\r)^{3/2}}\r]\nn\\
& &-\,\frac{2\,\pi}{3\,\beta^2}\,\f{\dot z}{1-{\dot z}^2},
\end{eqnarray}
with the first line representing the vacuum term~\cite{jaekel-1997,ford-1982} 
and the second line characterizing the finite temperature term.

\par

Let us emphasize here a few points regarding the radiation reaction force
that we have obtained above.
The procedure that we have adopted to arrive at the radiation reaction force 
is the same as the method that had been considered earlier (in this 
context, see Ref.~\cite{ford-1982}).
The earlier effort had arrived at the radiation reaction force in the 
quantum vacuum (\ie\/ at zero temperature), which matches with our result 
(provided a suitable Lorentz factor is accounted for).
As is well known, the radiation reaction force on the mirror in the quantum 
vacuum has exactly the same form as the radiation reaction force on a 
non-uniformly moving charge that one encounters in 
electromagnetism~\cite{ford-1982,jaekel-1997}. 
We should point out here that the procedure we have adopted and the complete
relativistic result we have obtained for the radiation reaction force is 
different from another prior effort in this direction (see 
Ref.~\cite{alves-2008}).
Nevertheless, we find that the results for the radiation reaction force match 
in the non-relativistic limit~\cite{fulling-1976,jaekel-1993a,alves-2008}, which 
is the domain of our primary interest in the latter part of this article.

\par

Until now, the results have been exact, and we have made no assumptions 
on the amplitude of the velocity of mirror.
When analyzing the Brownian motion of the mirror in the latter sections, 
we shall be working in the non-relativistic limit. 
In such a limit (\ie when $\vert {\dot z}\vert\ll 1$), the above mean
radiation reaction force simplifies to
\begin{equation}
\langle {\hat F}_{\rm rad}\rangle
=\frac{1}{6\,\pi}\,{\dddot z}
-\frac{2\,\pi}{3\,\beta^2}\,{\dot z},\label{eq:Frad-nr}
\end{equation}
where we have ignored factors of order ${\dot z}^2$.
Note that at large temperatures, it is the second term that proves to
be the dominant one.
The term describes the standard dissipative force proportional to the 
velocity that is expected to arise as a particle moves through a thermal 
bath.


\section{Correlation function describing the fluctuating 
component}\label{sec:fc}

As we discussed in the introductory section, apart from the dissipative
component, a particle moving through a thermal bath also experiences a  
fluctuating force.
We have evaluated the dissipative force on the moving mirror in the last
section. 
Let us now turn to the calculation of the correlation function that
governs the fluctuating component of the radiation reaction force.

\par

The fluctuating component of the force on the moving mirror is clearly
given by the deviations from the mean value.
The operator describing the random force on the mirror can be defined as
\begin{equation}
{\hat {\cal R}}(t)\equiv {\hat F}_{\rm rad}-\langle {\hat F}_{\rm rad}\rangle
=-\frac{\d {\hat P}^x}{\d t}+\frac{\d\langle \hat{P}^x\rangle}{\d t},
\label{eq:Frad-fc}
\end{equation}
where ${\hat P}_x$ is the momentum operator associated with the scalar 
field as given by Eq.~(\ref{eq:Px-op}).
Upon using the operator version of the conservation of the stress-energy
tensor, one can show that the random force acting on the moving mirror can
be expressed in terms of the components of the stress-energy tensor as
follows:
\begin{eqnarray}
{\hat {\cal R}}(t)
&=&-{\dot z}(t)\,\l[{\hat {\cal T}}^{01}_{_{\rm R}}(t,z)
-{\hat {\cal T}}^{01}_{_{\rm L}}(t,z)\r]\nonumber \\
& &+\,{\hat {\cal T}}^{00}_{_{\rm R}}(t,z)
-{\hat {\cal T}}^{00}_{_{\rm L}}(t,z),\label{eq:Rt}
\end{eqnarray}
where the quantity ${\hat {\cal T}}^{ab}(t,x)$ is defined as 
\begin{equation}
{\hat {\cal T}}^{ab}(t,x)
= \hat{T}^{ab}(t,x)-\langle \hat{T}^{ab}(t,x)\rangle.
\end{equation}
Therefore, the correlation function describing the fluctuating force
${\hat {\cal R}}(t)$ can be written as   
\begin{widetext}
\begin{eqnarray}
\langle {\hat {\cal R}}(t)\, {\hat {\cal R}}(t')\rangle
\!\!&=&\!\!{\dot z}\,{\dot z}'\,
\l[\langle {\hat {\cal T}}^{01}_{_{\rm R}}(t,z)\,
{\hat {\cal T}}^{01}_{_{\rm R}}(t',z')\rangle
+\langle {\hat {\cal T}}^{01}_{_{\rm L}}(t,z)\,
{\hat {\cal T}}^{01}_{_{\rm L}}(t',z')\rangle\r]
-{\dot z}\,\l[\langle {\hat {\cal T}}^{01}_{_{\rm R}}(t,z)\, 
{\hat {\cal T}}^{00}_{_{\rm R}}(t',z')\rangle
+\langle {\hat {\cal T}}^{01}_{_{\rm L}}(t,z)\,
{\hat {\cal T}}^{00}_{_{\rm L}}(t',z')\rangle\r]\nn\\
& &-\,{\dot z}'\, \l[\langle {\hat {\cal T}}^{00}_{_{\rm R}}(t,z)\,
{\hat {\cal T}}^{01}_{_{\rm R}}(t',z')\rangle
+\langle {\hat {\cal T}}^{00}_{_{\rm L}}(t,z)\,
{\hat {\cal T}}^{01}_{_{\rm L}}(t',z')\rangle\r]
+\langle {\hat {\cal T}}^{00}_{_{\rm R}}(t,z)\,
{\hat {\cal T}}^{00}_{_{\rm R}}(t',z')\rangle
+\langle {\hat {\cal T}}^{00}_{_{\rm L}}(t,z)\,
{\hat {\cal T}}^{00}_{_{\rm L}}(t',z')\rangle,\label{eq:RtRt'}\nn\\
\end{eqnarray}
where $z=z(t)$ and $z'=z(t')$.

\par

The quantity $\langle{\hat {\cal T}}^{ab}(t,x)\,
{\hat {\cal T}}^{cd}(t',x')\rangle$ is essentially the so-called noise 
kernel corresponding to the stress-energy tensor of the scalar field (in 
this context, see, for instance, Refs.~\cite{phillips-2001-03,cho-2015}).
Upon using the decomposition~(\ref{eq:phi-d}), the modes~(\ref{eq:m}) 
and the expressions~(\ref{eq:set}) for the stress-energy tensor, the 
noise kernel in the regions to the right and the left of the mirror 
can be calculated to be 
\begin{eqnarray}
\langle{\hat {\cal T}}^{ab}_{_{\rm R}}(t,x)\,
{\hat {\cal T}}^{cd}_{_{\rm R}}(t',x')\rangle
&=&\f{\pi^2}{8\,\beta^4}\,
\biggl\{(-1)^{a+b+c+d}\; {\rm cosech}^4\l[\pi\,(v-v')/\beta\r]
+(-1)^{a+b}\,p_1'^2(u')\; {\rm cosech}^4\l[\pi\,\l[v-p_1(u')\r]/\beta\r]\nn\\
& &+\,(-1)^{c+d}\, p_1'^2(u)\;
{\rm cosech}^4\l[\pi\,\l[p_1(u)-v'\r]/\beta\r]
+p_1'^2(u)\,p_1'^2(u')\;
{\rm cosech}^4\l[\pi\,\l[p_1(u)-p_2(u')\r]/\beta\r]\bigg\},\nn\\
\\
\langle{\hat {\cal T}}^{ab}_{_{\rm L}}(t,x)\,
{\hat {\cal T}}^{cd}_{_{\rm L}}(t',x')\rangle
&=&\f{\pi^2}{8\,\beta^4}\,\bigg\{{\rm cosech}^4\l[\pi\,(u-u')/\beta\r]
+(-1)^{c+d}\, p_2'^2(v')\, {\rm cosech}^4\l[\pi\,\l[u-p_2(v')\r]/\beta\r]\nn\\
& &+\,(-1)^{a+b}\,p_2'^2(v)\, {\rm cosech}^4\l[\pi\,\l[p_2(v)-u'\r]/\beta\r]\nn\\
& &+(-1)^{a+b+c+d}\,p_2'^2(v)\,p_2'^2(v')\,
{\rm cosech}^4\l[\pi\,\l[p_2(v)-p_2(v')\r]\beta\r]\biggr\},
\end{eqnarray}
where, as we had defined, $u=t-x$ and $v=t+x$, while $u'=t'-x'$ and 
$v'=t'+x'$.
Note that the indices $(a,b,c,d)$ take on the values zero and unity 
corresponding to $t$ and $x$, respectively. 
Along the trajectory of the mirror $z(t)$, the noise kernels to the 
right and the left of the mirror simplify to
\begin{subequations}
\label{eq:set-nk}
\begin{eqnarray}
\langle{\hat {\cal T}}^{ab}_{_{\rm R}}(t,z)\,
{\hat {\cal T}}^{cd}_{_{\rm R}}(t',z')\rangle
&=&\frac{\pi^2}{8\,\beta^4}\,\biggl[(-1)^{a+b+c+d}+(-1)^{a+b}\,
\l(\f{1+{\dot z}'}{1-\dot z'}\r)^2
+(-1)^{c+d}\, \l(\f{1+\dot z}{1-\dot z}\r)^2\nn\\
& &+\,\l(\frac{1+{\dot z}}{1-{\dot z}}\r)^2\,
\l(\frac{1+{\dot z}'}{1-{\dot z}'}\r)^2\biggr]\;
{\rm cosech}^4\l[\pi\,\l(\Delta t+\Delta z\r)/\beta\r],\label{eq:set-nkr}\\
\langle{\hat {\cal T}}^{ab}_{_{\rm L}}(t,z)\,
{\hat {\cal T}}^{cd}_{L}(t',z')\rangle
&=&\frac{\pi^2}{8\beta^4}\biggl[1+(-1)^{a+b}\,\l(\f{1-{\dot z}}{1+{\dot z}}\r)^2
+(-1)^{c+d}\,\l(\f{1-{\dot z}'}{1+{\dot z}'}\r)^2\nn\\
& &+\,(-1)^{a+b+c+d}\,\l(\f{1-{\dot z}}{1+{\dot z}}\r)^2\,
\l(\frac{1-{\dot z}'}{1+{\dot z}'}\r)^2\biggr]\;
{\rm cosech}^4\l[\pi\,\l(\Delta t-\Delta z\r)/\beta\r],\label{eq:set-nkl}
\end{eqnarray}
\end{subequations}
\end{widetext}
where $\Delta t = t-t'$ and $\Delta z= z-z'$. 
These quantities can be used in the expression~(\ref{eq:RtRt'}) to arrive
at the correlation function describing the fluctuating component of the
radiation reaction force.
Until now, the expressions we have obtained are exact.
Our aim is to arrive at the correlation function when the mirror is moving
non-relativistically.
If one consistently ignores terms of order ${\dot z}^2$, it can be shown that 
the correlation function simplifies to (for details, see App.~\ref{app:nrl-nk})
\begin{equation}
\langle{\hat {\cal R}}(t)\,{\hat {\cal R}}(t')\rangle
=\frac{\pi^2}{\beta^4}\,{\rm cosech}^4\l[\pi\,(t-t')/\beta\r].
\label{eq:RtRt'-fv}
\end{equation}
This correlation function is a sharply peaked function about $t=t'$ with a width
of the order of $\beta$. 
In the limit $\beta\to \infty$ (\ie in the quantum vacuum), this correlation 
function reduces to 
\begin{equation}
\langle{\hat {\cal R}}(t)\,{\hat {\cal R}}(t')\rangle
=\frac{1}{\pi^2}\,\f{1}{(t-t')^4},
\end{equation}
which is what can be expected from general arguments in $(1+1)$-spacetime
dimensions.
 

\section{Establishing the fluctuation-dissipation theorem}\label{sec:fdt}

Having obtained the average radiation reaction force on the moving mirror
and having evaluated the correlation function describing the fluctuating
component, let us now turn to establishing the fluctuation-dissipation
theorem relating these quantities.
In this section, we shall first explicitly establish the theorem for the
problem of the moving mirror in the frequency domain and then go on to 
also establish it in the time domain.


\subsection{The fluctuation-dissipation theorem in the frequency domain}

Fluctuation-dissipation theorem is a general result in statistical mechanics, 
which is a relation between the generalized resistance and the fluctuations 
of the generalised forces in linear dissipative systems~\cite{callen-1951,
kubo-1957,kubo-1966}. 
Before discussing the fluctuation-dissipation theorem let us define some 
essential quantities which are needed to state the fluctuation dissipation 
theorem. 

\par

Let us define the correlation function of an operator ${\hat A}$ as
\begin{equation}
C_A(t)\equiv\langle\hat A(t_0)\,\hat A(t_0+t)\rangle.
\end{equation}
The symmetric and anti-symmetric correlation functions, \ie\/ $C_A^+(t)$ and 
$C_A^-(t)$, of the operator ${\hat A}$ can be defined to be~\cite{kubo-1966}
\begin{subequations}
\begin{eqnarray}
C_A^+(t)
&\equiv&\f{1}{2}\,\l(\langle{\hat A}(t_0)\,{\hat A}(t_0+t)\rangle
+\langle{\hat A}(t_0+t)\,{\hat A}(t_0)\rangle\r)\label{csym}\nn\\
&=&\frac{1}{2}\,\l[C_A(t)+C_A(-t)\r],\\
C_A^-(t)
&\equiv&\f{1}{2}\,\l(\langle{\hat A}(t_0)\,{\hat A}(t_0+t)\rangle
-\langle{\hat A}(t_0+t)\,{\hat A}(t_0)\rangle\r)\label{cantsym}\nn\\
&=&\f{1}{2}\,\l[C_A(t)-C_A(-t)\r].
\end{eqnarray}
\end{subequations}
Given a function $f(t)$, let the Fourier transform ${\widetilde f}(\omega)$
be defined as
\begin{equation}
{\widetilde f}(\omega)
=\int_{-\infty}^\infty \d t\, f(t)\, {\rm e}^{-i\,\omega\, t}.
\end{equation}
The fluctuation-dissipation theorem describing the random function
${\hat A}(t)$ can be stated as the following relation between the 
Fourier transforms ${\widetilde C}_A^+(\omega)$ and 
${\widetilde C}_A^-(\omega)$~\cite{kubo-1966}: 
\begin{equation}
{\widetilde C}^+_A(\omega)
={\rm coth}(\beta\,\omega/2)\,{\widetilde C}^-_A(\omega),\label{eq:fdt1}
\end{equation}
with $\omega>0$.

\par

In the rest of this section, our aim will be to establish the 
relation~(\ref{eq:fdt1}) for the fluctuating component of the 
radiation reaction force on the moving mirror, 
\viz ${\hat {\cal R}}(t)$.
Note that the quantity $C_{\cal R}(t)$ can be written as [cf.
Eq.~(\ref{eq:RtRt'-fv})]
\begin{equation}
\label{corrcrt}
C_{\cal R}(t)
=\frac{\pi^2}{\beta^4}\,{\rm cosech}^4\l[\pi\,(t+i\,\epsilon)/\beta\r],
\end{equation}
where, as is usually done in the context of quantum field theory, we have
suitably introduced an $i\,\epsilon$ factor (with $\epsilon\to 0^{+}$) to 
regulate the two-point function in the coincidence limit.
The Fourier transform of the correlation function $C_{\cal R}(t)$ is, 
evidently, given by
\begin{equation}
{\widetilde C}_{\cal R}(\omega)
=\f{\pi^2}{\beta^4}\,
\int^\infty_{-\infty}\d t\,
{\rm cosech}^4\l[\pi\,(t+i\,\epsilon)/\beta\r]\,{\rm e}^{-i\,\omega\, t}.
\label{eq:CRomega-i}
\end{equation}
To evaluate this integral, it proves to be convenient to express the 
function $C_{\cal R}(t)$ as a series in the following fashion (for
details, see App.~\ref{app:CRt-sr}):
\begin{widetext}
\begin{eqnarray}
C_{\cal R}(t)
=\f{\pi^2}{\beta^4}\,{\rm cosech}^4\l[\pi\,(t+i\,\epsilon)/\beta\r]
&=&-\frac{2}{3\,\beta^2}\,
\l[\frac{1}{(t+i\,\epsilon)^2}
+\sum^\infty_{n=1}\frac{1}{(t+i\,n\,\beta)^2}
+\sum^\infty_{n=1}\frac{1}{(t-i\,n\,\beta)^2}\r]\nn\\
& &+\,\f{1}{\pi^2}\,
\l[\f{1}{(t+i\,\epsilon)^4}
+\sum^\infty_{n=1}\frac{1}{(t+i\,n\,\beta)^4}
+\sum^\infty_{n=1}\frac{1}{(t-i\,n\,\beta)^4}\r].\label{eq:RtRt'-sr}
\end{eqnarray}
\end{widetext}
Upon using this series representation, the integral~(\ref{eq:CRomega-i}) 
can be carried out as a contour integral in the complex $\omega$-plane.
Since $\omega>0$, the contour has to be closed in the lower half plane. 
The contour encloses the poles at $-i\,\epsilon$ and $-i\,n\,\beta$, so
that only the first two terms within the square brackets in the above
series representation for $C_{\cal R}(t)$ contribute.
Their contributions can be summed over to obtain that~\cite{jaekel-1993a,
jaekel-1993b,jaekel-1997,alves-2003,alves-2008} 
\begin{eqnarray}
{\widetilde C}_{\cal R}(\omega)
=\frac{2}{(1-{\rm e}^{-\beta\,\omega)}}
\l(\f{\omega^3}{6\,\pi}+\frac{2\,\pi\,\omega}{3\,\beta^2}\r).
\label{eq:CRomega}
\end{eqnarray}
The quantities $\widetilde{C}^+_R(\omega)$ and $\widetilde{C}^-_R(\omega)$ 
can be determined from the above expression for 
${\widetilde C}_{\cal R}(\omega)$, and they are found to be
\begin{subequations}
\begin{eqnarray}
{\widetilde C}_{\cal R}^+(\omega)
&=&{\rm coth}\l(\beta\omega/2\r)\,
\l(\f{\omega^3}{6\,\pi}+\f{2\,\pi\,\omega}{3\,\beta^2}\r),\\
{\widetilde C}_{\cal R}^-(\omega)
&=&\f{\omega^3}{6\,\pi}+\f{2\,\pi\,\omega}{3\,\beta^2}.
\end{eqnarray}
\end{subequations}
The first term in the above expression for ${\widetilde C}_{\cal R}^-(\omega)$
is the vacuum contribution, while the second term arises at a finite temperature.
These can be attributed to the ${\dddot z}$ and the ${\dot z}$ terms 
that arise in the mean radiation reaction force at zero and finite
temperature, respectively [cf. Eq.~(\ref{eq:Frad-nr})]. 
It is evident from these expressions that the quantities
$\widetilde{C}^+_R(\omega)$ and $\widetilde{C}^-_R(\omega)$ are 
related as
\begin{equation}
{\widetilde C}_{\cal R}^+(\omega)
={\rm coth}\l(\beta\,\omega/2\r)\, {\widetilde C}_R^-(\omega),
\end{equation}
exactly as required by the fluctuation-dissipation theorem.


\subsection{The fluctuation-dissipation theorem in the time domain}

Let us now consider the fluctuation-dissipation theorem in the time
domain.
In the time domain, the theorem relates the correlation function
$C_{\cal R}(t)$ of the fluctuating force to the amplitude of the
coefficient, say, $m\,\gamma$, of the mean dissipative force 
(proportional to velocity) arising at a finite temperature as 
follows~\cite{kubo-1966}:
\begin{equation}
m\,\gamma
= \beta\,\int_0^{\infty}\, \d t\, C_{\cal R}(t).
\end{equation}
In the case of the moving mirror, we have $m\,\gamma=2\,\pi/(3\,\beta^2)$
[cf.~Eq.~(\ref{eq:Frad-nr})].
Since the above integral corresponds to the $\omega 
\to 0$ of ${\widetilde C}_{\cal R}(\omega)/2$,
we find that 
\begin{equation}
\beta\,\int_0^{\infty}\, \d t\, C_{\cal R}(t)
=\beta\,\lim_{\omega\to 0}\f{{\widetilde C}_{\cal R}(\omega)}{2}
=\beta\,\frac{2\,\pi}{3\,\beta^3}
= m\,\gamma,
\end{equation}
as required, implying the validity of the fluctuation-dissipation theorem
in the time domain as well.


\section{Diffusion of the mirror at finite temperature}\label{sec:msd-ft}

In this section, we shall utilize the fluctuation-dissipation theorem to 
determine the mean-squared displacement in the position of the mirror due
to the combination of the mean radiation reaction force on the mirror as
well as the fluctuating component.
We shall also discuss the different limiting behavior of the mean-squared 
displacement of the mirror.

\par


\subsection{The mean-squared displacement of the mirror at finite temperature}

The mean-squared displacement $\sigma_z^2(t)$ in the position of the mirror is 
defined as
\begin{equation}
\sigma_z^2(t)\equiv \langle[{\hat z}(t)-{\hat z}(0)]^2\rangle
= 2\,\l[C_z^+(0)-C_z^+(t)\r],\label{eq:msd}
\end{equation}
where ${\hat z}(t)$ represents the stochastic nature of the position of the 
mirror, which are induced due to the fluctuations in the radiation reaction 
force.
When we take into account the mean radiation reaction force~(\ref{eq:Frad-nr})
and the fluctuating component~(\ref{eq:Frad-fc}), the Langevin equation governing
the motion of the moving mirror is given by
\begin{equation}
m\,\f{\d^2{\hat z}}{\d t^2}
-\f{1}{6\,\pi}\,\f{\d^3{\hat z}}{\d t^3}
+\f{2\,\pi}{3\,\beta^2}\,\f{\d{\hat z}}{\d t}
={\hat {\cal R}}(t).
\end{equation}
Let ${\widetilde z}(\omega)$ and ${\widetilde {\cal R}}(\omega)$ denote
the Fourier transforms of the position of the mirror ${\hat z}(t)$ and the
fluctuating component ${\hat {\cal R}}(t)$ of the radiation reaction force.
The above Langevin equation relates these two quantities as follows:
\begin{equation}
{\widetilde z}(\omega)
=\widetilde{\chi}(\omega)~\widetilde{\cal{R}}(\omega),\label{eq:cs}
\end{equation}
where ${\widetilde \chi}(\omega)$ is a complex quantity known as the 
generalized susceptibility.
It can be expressed as
\begin{equation}
\label{compsus}
{\widetilde \chi}(\omega)
=\f{6\,\pi}{i\,\omega\,(\omega+i\,\alpha_1)\,
(\omega-i\,\alpha_2)}
\end{equation}
with $\alpha_1$ and $\alpha_2$ being given by 
\begin{subequations}
\label{eq:alpha}
\begin{eqnarray}
\alpha_1&=& 3\,\pi\, \omega_{\rm c}\, 
\l[\sqrt{1+\l(\f{2\,r}{3}\r)^2}+1\r],\\
\alpha_2&=& 3\,\pi\, \omega_{\rm c}\, 
\l[\sqrt{1+\l(\f{2\,r}{3}\r)^2}-1\r],
\end{eqnarray}
\end{subequations}
where we have set $\omega_{\rm c}= m$ and $r=(\beta\,m)^{-1}$.
The quantity $\omega_{\rm c}$ is essentially the Compton frequency associated
with the mirror, while $r$ is the dimensionless ratio of the average energy 
associated with a single degree of freedom in the thermal bath and the rest
mass energy of the mirror.

\par

Let us write the generalized susceptibility as $\chi(\omega)
={\widetilde \chi}'(\omega)-i\,{\widetilde \chi''}(\omega)$, where 
${\widetilde \chi}'(\omega)$ and ${\widetilde \chi}''(\omega)$ are
real quantities. (The single and the double primes above 
${\tilde \chi}(\omega)$ are the conventional notations to denote the 
real and the imaginary parts of the generalized susceptibility. 
It should be clarified that these primes {\it do not}\/ represent 
derivatives of these quantities.)
According to the fluctuation-dissipation theorem, the quantity 
${\widetilde C}_z^+(\omega)$ is related to the quantity 
${\widetilde \chi}''(\omega)$ as follows~\cite{kubo-1966}:
\begin{equation}
{\widetilde C}_z^+(\omega)
={\rm coth}\l(\beta\,\omega/2\r)\,\widetilde\chi''(\omega).
\label{eq:fdt}
\end{equation}
Note that the mean-squared displacement $\sigma_z^2(t)$ of the mirror is
related to the correlation function $C_z^+(t)$ [cf. Eq.~(\ref{eq:msd})].
The correlation function $C_z^+(t)$ can be arrived at by inverse Fourier 
transforming the above expression for ${\widetilde C}_z^+(\omega)$.
Clearly, the quantity $C_z^+(t)$ is the convolution of the inverse 
Fourier transforms of ${\rm coth}\l(\beta\,\omega/2\r)$ and 
${\widetilde \chi}''(\omega)$, so that we have
\begin{equation}
C_z^+(t)=\frac{i}{\beta}\,\int_{-\infty}^{\infty}\,\d t'\,
{\rm coth}\l(\pi\,t'/\beta\r)\,\chi''(t-t'),\label{eq:msd-c}
\end{equation}
where $\chi''(t)$ is described by the integral
\begin{equation}
\chi''(t)=\f{1}{2\,\pi}\,\int_{-\infty}^{\infty}\d \omega\,
{\widetilde \chi}''(\omega)\,{\rm e}^{i\,\omega\,t}.
\label{eq:chi-t}
\end{equation}

\par

The imaginary part of complex susceptibility ${\widetilde \chi}(\omega)$ 
is found to be
\begin{equation}
{\widetilde \chi}''(\omega)
=\f{6\,\pi\,(\omega^2+\alpha_1\,\alpha_2)}{(\omega-i\,\epsilon)\,
(\omega^2+\alpha_1^2)\,(\omega^2+\alpha_2^2)},\label{eq:ip-chi}
\end{equation}
where we have introduced an $i\,\epsilon$ factor suitably to ensure the 
convergence of ${\tilde \chi}(\omega)$~\cite{kubo-1957}. 
The integral~(\ref{eq:chi-t}), with ${\widetilde \chi}''(\omega)$ given by 
the above expression, can be carried out easily as a contour integral in 
the complex $\omega$-plane, and one obtains that  
\begin{widetext}
\begin{equation}
\label{chidpt}
\chi''(t)=3\,i\,\pi\,
\biggl\{\frac{2\,\Theta(t)}{\alpha_1\,\alpha_2}
-{\rm sgn}(t)\,\l[\f{{\rm e}^{-\alpha_1\,|t|}}{\alpha_1\,(\alpha_1+\alpha_2)}
+\f{{\rm e}^{-\alpha_2\,|t|}}{\alpha_2\,(\alpha_1+\alpha_2)}\r]\biggr\},
\end{equation}
where $\Theta(t)$ is the theta function, while the function $\rm{sgn(t)}$ 
is given by
\begin{equation}
{\rm sgn}(t)=\l\{\begin{array}{ll}
1 &{\rm when}\quad t>0,\\
-1\, &{\rm when}\quad t<0.                        
\end{array}\r.\label{eq:sgnt}
\end{equation}
Upon using the above expression for $\chi''(t)$ in Eq.~(\ref{eq:msd-c}), we 
find that we can write $C_z^+(t)$ as follows:
\begin{eqnarray}
C_z^+(t)
&=&-\f{6\,\pi}{\alpha_1\,\alpha_2\,\beta}\, 
\int^t_{-\infty}\,\d t'\,{\rm coth}\l[\pi\,(t'+i\epsilon)/\beta\r]
+\frac{3\,\pi}{\alpha_1\,(\alpha_1+\alpha_2)\,\beta}\,
\l[{\rm e}^{-\alpha_1\,t}\,I_1(\alpha_1,t)
-{\rm e}^{\alpha_1\,t}\,I_2(\alpha_1,t)\r]\nn\\
& &+\,\frac{3\,\pi}{\alpha_2\,(\alpha_1+\alpha_2)\,\beta}\,
\l[{\rm e}^{-\alpha_2\,t}\,I_1(\alpha_2,t)-{\rm e}^{\alpha_2\,t}\,
I_2(\alpha_2,t)\r],
\end{eqnarray}
where the quantities $I_1(\alpha,t)$ and $I_2(\alpha,t)$ are described by
the integrals
\begin{subequations}
\label{eq:I-at}
\begin{eqnarray}
I_1(\alpha,t)
&=&\int_{-\infty}^t\,\d t'\,
{\rm e}^{\alpha\, t'}\,{\rm coth}\l[\pi\,(t'+i\,\epsilon)/\beta\r],\\
I_2(\alpha,t)
&=&\int_{t}^\infty\,\d t'\, {\rm e}^{-\alpha\, t'}\,
{\rm coth}\l[\pi\,(t'+i\epsilon)/\beta\r].
\end{eqnarray}
\end{subequations}
On substituting the above expression for $C_z^+(t)$ in Eq.~(\ref{eq:msd}), 
we obtain the mean-squared displacement of the mirror to be
\begin{eqnarray}
\sigma_{z}^2(t)
&=&\frac{12\,\pi}{\alpha_1\alpha_2\,\beta}\,
\int^t_{0}\d t'\,{\rm coth}\l[\pi\,(t'+i\epsilon)/\beta\r]
-\frac{6\,\pi}{\alpha_1\,(\alpha_1+\alpha_2)\,\beta}\,
\l[{\rm e}^{-\alpha_1\,t}\,I_1(\alpha_1,t)
-{\rm e}^{\alpha_1\,t}\,I_2(\alpha_1,t)
-I_1(\alpha_1,0)+I_2(\alpha_1,0)\r]\nn\\ \
& &-\,\f{6\,\pi}{\alpha_2\,(\alpha_1+\alpha_2)\,\beta}\,
\l[{\rm e}^{-\alpha_2\,t}\,I_1(\alpha_2,t)
-{\rm e}^{\alpha_2\,t}\,I_2(\alpha_2,t)
-I_1(\alpha_2,0)+I_2(\alpha_2,0)\r].\label{eq:msd-pv}
\end{eqnarray}
The integrals $I_1(\alpha,t)$ and $I_2(\alpha,t)$ can be evaluated in terms 
of the hypergeometric functions (for details, see App.~\ref{app:ei}), and
the final result can be expressed as
\begin{eqnarray}
\sigma_{z}^2(t)
=\f{12}{\alpha_1\,\alpha_2}\,
\biggl\{\gamma_{\rm E}+{\rm ln}\l[2\,{\rm sinh}(\pi\,t/\beta\r)]\biggr\}
+\f{12}{\alpha_1\,(\alpha_1+\alpha_2)}\,F(p_1,t)
+\f{12}{\alpha_2\,(\alpha_1+\alpha_2)}\,F(p_2,t),\label{eq:msd-ft-fv}
\end{eqnarray}
where $\gamma_{_{\rm E}}\simeq 0.5772$ is the Euler-Mascheroni 
constant~\cite{gradshteyn-2007}.
The function $F(p,t)$ is given by
\begin{eqnarray}
F(p,t)
=\frac{\pi}{2}\, {\rm cot}(\pi\, p)\; {\rm e}^{-2\,\pi\,p\, t/\beta} 
+\frac{{\rm e}^{-2\,\pi\,t/\beta}}{2\,(1-p)}\,
{}_2F_1\l[1,1-p;2-p;{\rm e}^{-2\,\pi\,t/\beta}\r]\,
+\frac{1}{2\,p}\, {}_2F_1\l[1,p;p+1;{\rm e}^{-2\pi\,t/\beta}\r]+\psi_0(p),
\quad
\end{eqnarray}
\end{widetext}
where ${}_2F_1[a,b,c,;z]$ denotes the hypergeometric function, and
$\psi_n(z)$ is known as the polygamma function~\cite{gradshteyn-2007}.
The quantities $p_1$ and $p_2$ are defined as 
\begin{equation}
p_1 =\f{\alpha_1\,\beta}{2\,\pi},\quad
p_2 =\f{\alpha_2\,\beta}{2\,\pi},\label{eq:p1-p2}
\end{equation}
with $\alpha_1$ and $\alpha_2$ being given by Eqs.~(\ref{eq:alpha}).
Note that the mean-squared displacement~(\ref{eq:msd-ft-fv}) depends on three 
time scales, \viz\/ $t$, $\omega_{\rm c}^{-1}$ and $\beta$. 
Let us now consider the limiting forms of the mean-squared displacement 
of the mirror in the different regimes of interest.


\subsection{The different limiting behavior of the mean-squared displacement 
of the mirror}

As mentioned above, the mean-squared displacement $\sigma_{z}^2(t)$ depends 
on three time scales $t$, $\omega_c^{-1}$ and $\beta$. 
Using these time scales one can construct the following three dimensionless 
variables: $\omega_{\rm c}\, t$, $t/\beta$ and $\beta\,\omega_c$. 
Notice that the expression~(\ref{eq:msd-ft-fv}) for the mean-squared displacement 
at a finite temperature depends on time only through the following dimensionless 
combination: ${\tilde t}\equiv t/\beta$.
It also depends on the dimensionless quantity $r=(\beta\,\omega_{\rm c})^{-1}$, 
which we had introduced earlier [cf. Eq.~(\ref{eq:alpha})].
Typically, we will be interested in the behavior of the mean-squared
displacement at small and large times, \ie\/ for ${\tilde t} \ll 1$ 
and ${\tilde t}\gg 1$.
But, because of the presence of the additional dimensionless quantity $r$, 
the different possible limits that one can actually consider are as follows:
\begin{eqnarray}
\lim\limits_{r\to 0}\,\lim\limits_{{\tilde t}\to 0}\,\sigma_{z}^2(t),&
\lim\limits_{{\tilde t}\to 0}\,\lim\limits_{r\to 0}\,\sigma_{z}^2(t),\nn\\
\lim\limits_{r\to \infty}\,\lim\limits_{{\tilde t}\to 0}\,\sigma_{z}^2(t),& 
\lim\limits_{{\tilde t}\to 0}\,\lim\limits_{r\to \infty}\,\sigma_{z}^2(t),\nn
\end{eqnarray}
for small ${\tilde t}$, and
\begin{eqnarray}
\lim\limits_{r\to 0}\,\lim\limits_{{\tilde t}\to \infty}\,\sigma_{z}^2(t),& 
\lim\limits_{{\tilde t}\to \infty}\,\lim\limits_{r\to 0}\,\sigma_{z}^2(t),\nn\\
\lim\limits_{r\to \infty}\,\lim_{{\tilde t}\to \infty}\,\sigma_{z}^2(t),& 
\lim\limits_{{\tilde t}\to \infty}\,\lim\limits_{r\to \infty}\,\sigma_{z}^2(t),\nn
\end{eqnarray}
for large ${\tilde t}$.
\enlargethispage*{1.0cm}
In other words, apriori, one can consider the limits of small and large $r$ 
before or after considering the small and large limits of ${\tilde t}$.
However, we find that, as $r\to 0$ (or, as $r\to \infty$) the limiting values 
of the mean-squared displacement are not numerically equal to the dominant 
term in the series expansion of $\sigma_{z}^2(t)$ around $r=0$ (and 
$r=\infty$, respectively) for all values of ${\tilde t}$. 
Therefore, we shall take the small and large limits of ${\tilde t}$, before 
considering the limiting cases of $r$. 

\par

We find that, in the limit of ${\tilde t}\ll 1$, $\sigma_{z}^2(t)$ can be
expressed as
\begin{eqnarray}
\sigma_z^2(t)
&=& 6\,\,t^2\,\biggl\{\f{3}{2}-\gamma_{\rm E}
-{\rm ln}\l(2\,\pi\, t/\beta\r)\nn \\
& &-\,\frac{1}{(p_1+p_2)}\,\l[1+p_1\,\psi_{0}(p_1)+p_2\,\psi_{0}(p_2)\r]\biggr\}.
\qquad\label{sigmass}
\end{eqnarray}
Whereas, when ${\tilde t}\gg 1$, it reduces to
\begin{eqnarray}
\sigma_z^2(t)
&=&\frac{3\,\beta\,t}{\pi}+\frac{6\,\beta^2}{2\,\pi^2}\,
\biggl\{\gamma_{\rm E}
+\f{p_2}{(p_1+p_2)}\,\l[\f{1}{2\,p_1}+\psi_0(p_1)\r]\nn\\
& &+\,\frac{p_1}{(p_1+p_2)}\,\l[\frac{1}{2\,p_2}+\psi_0(p_2)\r]\biggr\}.
\label{sigmal}
\end{eqnarray}
Let us now consider the different limits of $r$ of the above two expressions.
For convenience and clarity, we have listed these forms in the table below
and have commented appropriately on their behavior.
\begin{widetext}
\begin{center}
\begin{tabular}{| c | >{\centering\arraybackslash}m{5cm} | p{10cm} |}
\hline
& &\\
Relevant limits & Limiting behavior of $\sigma_{z}^2(t)$ 
& Remarks \hfill \\ 
& &\\
\hline
\hline
& & \multirow{7}{10cm}{Although we quote it for the sake of completeness, this 
limit corresponds to $\omega_{\rm c}\,t\ll 1$, \ie\/ when the times involved 
are much smaller than the Compton time scale.
The quantum nature of the mirror cannot be ignored in such a domain. 
Since our analysis assumes a classical, non-relativistic description for the 
mirror, it might be unjustified to attach any significance to this limit for
the mean-squared displacement of the mirror.}\\
$t\ll\omega_{\rm c}^{-1}\ll\beta$ 
& $6\, t^2\, \l[(3/2)-⁠\gamma_{\rm E}-⁠\ln (6\,\pi\, \omega_c\, t)\r]$ & \\ 
& &\\
\cline{1-2}
& &\\
$t\ll\beta\ll\omega_{\rm c}^{-⁠1}$ & $6\,t^2\,\l[1-⁠\ln\l(2\pi\,t/\beta\r)\r]$ & \\
& &\\
& &\\
\hline
\hline
& & \multirow{9}{10cm}{This limit demonstrates that, as long as $t \gg \beta$, 
the limiting behaviour of $\sigma_z^2(t)$ does not depend on $\omega_{\rm c}\, t$ 
(although the same comment as above applies to the case $\omega_{\rm c}\, t \ll 1$). 
Moreover, as is evident from the expression in the last row (below), this limit 
is also independent of $r$. 
\vskip 4pt
One can therefore see that, for $t \gg \beta$, the mirror exhibits the standard 
random walk with $\sigma_z^2(t) \propto t$.
(To highlight this behavior, we have expressed the final result in terms of the
parameter $\gamma$ to facilitate comparison with standard discussions of random 
walk~\cite{reif-1965}.)}\\
& &\\
$\beta \ll t \ll \omega_{\rm c}^{-⁠1}$ & &\\
& \multirow{2}{5cm}{\begin{equation*}
\f{2\, t}{m\,\gamma\,\beta}
+\f{3\,\beta^2}{2\,\pi^2}\simeq \f{2\,t}{m\,\gamma\,\beta}
\end{equation*}} & \\
& &\\
& &\\
\cline{1-1}
& &\\
& &\\
$\beta \ll \omega_{\rm c}^{-1}\ll t$ & &\\
& &\\
& &\\
\hline
\hline
& & \multirow{4}{10cm}{In these limits, the mirror behaves exactly like a 
Brownian particle.
\vskip 4pt
For $t \ll \beta$, we have $\sigma_z^2(t) \propto t^2$, and the mirror diffuses 
like a free particle with velocity $1/\sqrt{m\,\beta}$. 
This result suggests that the thermal length scale ($\beta$) can be the mean 
free path of the mirror. 
For $t \gg \beta$, we recover the standard random walk result, \viz\/ $\sigma_z^2(t) 
\propto t$.}\\
$\omega_{\rm c}^{-⁠1}\ll t\ll\beta$ &
$t^2/(\beta\, m)$ & \\
& &\\
\cline{1-2}
$\omega_{\rm c}^{-⁠1}\ll \beta\ll t$ &
\begin{equation*}
\f{2}{m\,\gamma\,\beta}\,\l[t-⁠\gamma^{-1}\r]\simeq \f{2\,t}{m\,\gamma\,\beta}
\end{equation*} & \\
\hline
\end{tabular}
\end{center}
\end{widetext}

\section{Diffusion of the mirror at zero temperature}\label{sec:msd-zt}

Let us now study the nature of diffusion of the mirror at zero temperature.


\subsection{The mean-squared displacement of the mirror at zero temperature}

At zero temperature, evidently, the finite temperature contribution will be
absent and the Langevin equation governing the motion of the mirror 
simplifies to
\begin{equation}
\label{let0}
m\,\f{\d^2{\hat z}}{\d t^2}
-\f{1}{6\,\pi}\,\f{\d^3{\hat z}}{\d t^3}={\hat {\cal R}}.
\end{equation} 
In such a case, the complex susceptibility ${\widetilde \chi}(\omega)$ is 
given by [cf.~Eq.~(\ref{eq:cs})]
\begin{equation}
\label{chiomega}
{\widetilde \chi}(\omega)
=\frac{6\,\pi}{i\,\omega^2\,(\omega+6\,\pi\,i\,\omega_{\rm c})},
\end{equation}
and the imaginary part of the complex susceptibility $\widetilde{\chi}(\omega)$ 
can be determined to be
\begin{equation}
{\widetilde \chi}''(\omega)
=\frac{6\,\pi}{\omega\,[\omega^2+(6\,\pi\,\omega_{\rm c})^2]}.
\label{eq:ip-cs-zt}
\end{equation}

\par

At zero temperature, the fluctuation-dissipation relation~(\ref{eq:fdt}) 
reduces to
\begin{equation}
{\widetilde C}_z^+(\omega)
= [\Theta(\omega)-\Theta(-\omega)]\,{\widetilde \chi}''(\omega),
\label{eq:zt-fdt-fd}
\end{equation}
where $\Theta(\omega)$ denotes the theta function.
The inverse Fourier transform of this function yields $C^+_z(t)$, 
which is, evidently, a convolution described by the integral
\begin{equation}
\label{czt}
C_{z}^+(t)=\f{i}{\pi}\,
\int_{-\infty}^\infty\,\frac{\d t'}{t'}\,\chi''(t-t').
\end{equation}
The quantity $\chi''(t)$ can be easily evaluated from 
${\widetilde \chi}''(\omega)$ above [cf.~Eq.~(\ref{eq:ip-cs-zt})] 
as a contour integral in the complex $\omega$-plane.
It can be obtained to be
\begin{eqnarray}
\label{chit}
\chi''(t)
=\f{3\,\pi\, i}{(6\,\pi\,\omega_c)^2}\,
\l[2\,\Theta(t)-{\rm sgn}(t)\,{\rm e}^{-6\,\pi\,\omega_{\rm c}\,|t|}\r],
\end{eqnarray}
where ${\rm sgn}(t)$ is defined in Eq.~(\ref{eq:sgnt}).
On using this expression, we find that $ C_{z}^+(t)$ can be written as
\begin{eqnarray}
\label{cztf}
C_z^+(t)
&=&\f{-3}{(6\,\pi\,\omega_{\rm c})^2}\,
\biggl[2\,\int^t_{-\infty}\f{dt'}{t'}
-{\rm e}^{-6\,\pi\,\omega_{\rm c}\,t}\,
Ei\l(6\,\pi\,\omega_{\rm c}\,t\r)\nn\\
& &-\,{\rm e}^{6\,\pi\,\omega_{\rm c}\,t}\,
Ei\l(-6\,\pi\,\omega_{\rm c}\,t\r)\biggr],
\end{eqnarray}
where $Ei(x)$ is the exponential integral function~\cite{gradshteyn-2007}.
Upon using the above result, one can show that the mean-squared displacement 
of the mirror can be expressed as follows:
\begin{eqnarray}
\sigma_z^2(t)
&=&\frac{6}{(6\,\pi\,\omega_{\rm c})^2}\,
\biggl[2\,{\rm ln}(6\,\pi\,\omega_{\rm c}\,t)+2\,\gamma_E\nn\\
& &-\,{\rm e}^{-6\,\pi\,\omega_{\rm c}\,t}\,
Ei(6\,\pi\,\omega_{\rm c}\,t)\nn\\
& &-\,{\rm e}^{6\,\pi\,\omega_{\rm c}\,t}\,
Ei(-6\,\pi\,\omega_{\rm c}\,t)\biggr],
\end{eqnarray}
where, as we have pointed out before, $\gamma_{_{\rm E}}$ is Euler-Mascheroni 
constant.
                                              

\subsection{The different limiting behavior of the mean-squared displacement 
of the mirror}

We find that, when $\omega_{\rm c}\,t\ll 1$, the mean-squared displacement 
of the mirror behaves as
\begin{equation}
\sigma_{z}^2(t)
=6\,t^2\,\l[\frac{3}{2}-\gamma_{_{\rm E}}
-{\rm ln}\l(6\,\pi\, \omega_{\rm c}\, t\r)\r].
\end{equation}
Whereas, when $\omega_{\rm c}\,t\gg 1$, $\sigma_z^2(t)$ is found to behave 
as
\begin{equation}
\label{simat0}
\sigma_z^2(t)
=\f{12}{(6\,\pi\,\omega_{\rm c})^2}\,
\l[\gamma_{_{\rm E}}+{\rm ln}\l(6\,\pi\,\omega_{\rm c}\,t\r)\r].
\end{equation}
This implies that, at zero temperature, the mirror diffuses logarithmically
rather than linearly as it does at a finite temperature.
It should be mentioned that such a logarithmic diffusive behavior has been
arrived at earlier~\cite{jaekel-1993a} and it seems to be a general
characteristic of Brownian motion at zero temperature (in this context, see
Ref.~\cite{sinha-1992}).


\section{Discussion}\label{sec:d}

In this work, we have studied the random motion of a mirror that is 
immersed in a thermal bath.
We have explicitly evaluated the correlation function describing the
fluctuating component of the radiation reaction force on the moving
mirror and have established the fluctuation-dissipation theorem
relating the correlation function to the amplitude of the finite
temperature contribution to the radiation reaction force.
Also, utilizing the fluctuation-dissipation theorem, we have calculated
the mean-squared displacement of the moving mirror both at a finite as
well as at zero temperature.
We should stress that, in contrast to the earlier efforts, we have been 
able to arrive at a complete expression for the mean-squared displacement 
of the mirror that is valid at all times.
While we recover the standard results in the required limits at finite 
temperature, interestingly, we find that the mirror diffuses logarithmically 
at zero temperature, a result which confirms similar conclusions that have 
been arrived at earlier. 

\par

Finally, we find that the mean-squared displacement in the quantum vacuum 
cannot be obtained by blindly considering the zero temperature limit of the 
final expression for the mean-squared displacement at finite temperature.  
This is essentially because of the following reason: the integral representations
leading to the hypergeometric functions that arise in the finite temperature case 
[cf. Eq.~(\ref{eq:msd-ft-fv})] do not apply at zero temperature, thereby rendering 
the subsequent expressions invalid in this limit. 
(We have discussed this issue more quantitatively in App.~\ref{app:div}.) 
It is for this reason that, to analyze the zero temperature case, we have returned
to the Langevin equation and then proceeded with the derivation by making use of the  
corresponding fluctuation-dissipation theorem [see Eq.~(\ref{eq:zt-fdt-fd})].


\section{Acknowledgements}

D.J.S. thanks the Indian Institute of Technology Madras, Chennai, India, for the 
financial support through Half Time Research Assistantship. D.K. acknowledges support 
from the Department of Science and Technology (DST), India, through it’s INSPIRE Faculty 
Award. L.S. also wishes to thank the Indian Institute of Technology Madras, Chennai,
India, for the support through the New Faculty Seed Grant.


\appendix

\begin{widetext}
\section{Non-relativistic limit of the noise kernels}\label{app:nrl-nk}

In this appendix, we shall provide a few essential steps concerning the 
evaluation of the correlation function describing the fluctuating 
component of the radiation reaction force on the moving mirror.
 
\par

Note that we are interested in the correlation function when the mirror is
moving non-relativistically, \ie when $\vert {\dot z}\vert \ll 1$.
In such a limit, the noise-kernels~(\ref{eq:set-nk}) reduce to 
\begin{subequations}
\begin{eqnarray}
\langle\hat{\mathcal T}^{ab}_{_{\rm R}}(t,z)\,
\hat{\mathcal T}^{cd}_{_{\rm R}}(t',z')\rangle 
&\simeq&\frac{\pi^2}{8\,\beta^4}\,
\biggl[(-1)^{a+b+c+d}+(-1)^{a+b}\,\l(1+4\,{\dot z}'\r)
+(-1)^{c+d}\,\l(1+4\,{\dot z}\r)
+\l(1+4\,{\dot z}\r)\,\l(1+4\,{\dot z}'\r)\biggr]\nn\\
& &\times\,{\rm cosech}^4\l[\pi\,(\Delta t+\Delta z)/\beta\r],\label{nrnkr}\\
\langle\hat{\mathcal T}^{ab}_{_{\rm L}}(t,z)\,
\hat{\mathcal T}^{cd}_{_{\rm L}}(t',z')\rangle 
&\simeq&\f{\pi^2}{8\,\beta^4}\,
\biggl[1+(-1)^{a+b}\,\l(1-4\,{\dot z}\r)
+(-1)^{c+d}\, \l(1-4\,{\dot z}'\r)
+(-1)^{a+b+c+d}\,\l(1-4\,{\dot z}\r)\,\l(1-4\,{\dot z}'\r)\biggr]\nn\\
& &\times\,{\rm cosech}^4\l[\pi\,(\Delta t-\Delta z)/\beta\r].\label{nrnkl}
\end{eqnarray}
\end{subequations}
Upon substituting these results in the expression~(\ref{eq:RtRt'}), we 
get
\begin{eqnarray}
\langle{\hat {\cal R}}(t)\,{\hat {\cal R}}(t')\rangle 
&\simeq&\f{\pi^2}{2\,\beta^4}\,
\biggl\{{\rm cosech}^4\l[\pi\,\l(\Delta t-\Delta z\r)/\beta\r]
+{\rm cosech}^4\l[\pi\,\l(\Delta t+\Delta z\r)/\beta\r]\biggr\}\nn\\
& &-\,\f{\pi^2}{\beta^4}\,\l({\dot z}+{\dot z}'\r)\,
\biggl\{{\rm cosech}^4\l[\pi\,\l(\Delta t-\Delta z\r)/\beta\r]
-{\rm cosech}^4\l[\pi\,\l(\Delta t+\Delta z\r)/\beta\r]\biggr\}.
\label{eq:RtRt'-pv}
\end{eqnarray}
Using the series representation of ${\rm cosech}^4z$ 
[cf. Eq.~(\ref{eq:RtRt'-sr}); also see App.~\ref{app:CRt-sr}], we can
write 
\begin{eqnarray}
{\rm cosech}^4\l[\pi\,\l(\Delta t\pm \Delta z\r)/\beta\r]
&=&{\rm cosech}^4\l(\pi\,\Delta t/\beta\r)\nn\\
& &\pm\, \f{\Delta z}{\Delta t}\,
\l\{\f{4}{3\,\pi^2}\,\l(\f{\beta}{\Delta t}\r)^2\,
\sum_{n=-\infty}^{\infty}\frac{1}{\l[1+(i\,n\,\beta/\Delta t)\r]^3}
-\frac{4}{\pi^4}\,\l(\frac{\beta}{\Delta t}\r)^4\,
\sum_{n=-\infty}^{\infty}\frac{1}{\l[1+(i\,n\,\beta/\Delta t)\r]^5}\r\}\nn\\
\label{eq:sr-sv}
\end{eqnarray}
\end{widetext}
and, if we now make use of Eq.~(\ref{eq:sr-sv}) in (\ref{eq:RtRt'-pv}), we 
finally arrive at
\begin{equation}
\langle{\hat {\cal R}}(t)\,{\hat {\cal R}}(t')\rangle
=\frac{\pi^2}{\beta^4}\,{\rm cosech}^4\l(\pi\,\Delta t/\beta\r),
\end{equation}
which is the result we have quoted.


\section{Series representation of the correlation function}\label{app:CRt-sr}

In this appendix, we shall outline the method to arrive at the series 
representation~(\ref{eq:RtRt'-sr}) for the function $C_{\cal R}(t)$.

\par
  
We shall make use of the polygamma function $\psi_n(z)$ to arrive at the 
series representation for ${\rm cosech}^4(z)$. 
The polygamma function is defined as~\cite{gradshteyn-2007}
\begin{equation}
\psi_n(z)= \f{\d^{n+1}}{\d z^{n+1}}\ln \Gamma(z),
\end{equation}
where $\Gamma(z)$ is the gamma function.
The function $\psi_n(z)$ can be represented as an integral as follows:
\begin{equation}
\psi_n(z)=(-1)^{n+1}\int_0^{\infty}\d t\, 
\frac{{\rm e}^{-z\, t}\,t^n}{1-{\rm e}^{-t}}.
\end{equation}
Using this expression, we can write
\begin{equation}
\label{psi1}
\psi_1(i\,z)+\psi_1(-i\,z)
=2\,\int_0^{\infty}\,\d t\,\f{t\,{\rm cos}(z\,t)}{1-{\rm e}^{-t}}
\end{equation}
and, upon expressing ${\rm cos}(z\,t)$ and $(1-{\rm e}^{-t})^{-1}$ as 
a power series, we obtain that
\begin{eqnarray}
\psi_1(i\,z)+\psi_1(-i\,z)
&=&2\,\sum_{m=0}^{\infty}\f{(-1)^m\,z^{2\,m}}{(2\,m)!}\nn\\
& &\times\,\sum_{n=0}^{\infty} 
\int_0^{\infty}\d t\,{\rm e}^{-n\,t}\,t^{2\,m+1}.\qquad
\end{eqnarray}
Evaluating the integral, one obtains
\begin{eqnarray}
\psi_1(iz)+\psi_1(-iz)
&=&2\,\sum_{n=0}^{\infty}\f{1}{n^2}
\sum_{m=0}^{\infty}\biggl[(-1)^m\l(\f{z}{n}\r)^{2\,m}\nn\\
& &+\,2\,(-1)^m\, m\, \l(\f{z}{n}\r)^{2\,m}\biggr],
\end{eqnarray}
and carrying the sum over $m$ leads to
\begin{equation}
\psi_1(i\,z)+\psi_1(-i\,z)
=-\sum_{n=0}^{\infty}
\l[\f{1}{(z+i\,n)^2}+\frac{1}{(z-i\,n)^2}\r].
\end{equation}
The above series can easily be summed to arrive at~\cite{gradshteyn-2007} 
\begin{equation}
\sum_{n=0}^{\infty}
\l[\frac{1}{(z+i\,n)^2}+\frac{1}{(z-i\,n)^2}\r]
=\frac{1}{z^2}+\pi^2 {\rm cosech}^2(\pi\, z),
\end{equation}
so that we have
\begin{equation}
\psi_1(i\,z)+\psi_1(-i\,z)=-\frac{1}{z^2}-\pi^2\, {\rm cosech}^2(\pi z).
\label{eq:psi1-f}
\end{equation}

\par

\begin{widetext}
From the definition of polygamma function it is clear that $\psi_3(z)
=\d^2\psi_1(z)/\d z^2$. 
Upon substituting the result~(\ref{eq:psi1-f}) in this identity, we 
obtain that
\begin{equation}
\psi_3(i\,z)+\psi_3(-i\,z)
=\frac{6}{z^4} +4\,\pi^4\, {\rm cosech}^2(\pi\, z)
+ 6\,\pi^4\, {\rm cosech}^4(\pi\, z).\label{eq:psi3}
\end{equation}
From the integral representation of $\psi_n(z)$, we have
\begin{equation}
\psi_3(i\,z)+\psi_3(-i\,z)
=2\,\int_0^{\infty}\d t\,\frac{t^3\,{\rm cos}(z\,t)}{1-{\rm e}^{-t}}
=2\,\sum_{m=0}^{\infty}(-1)^m\,\f{z^{2\,m}}{(2\,m)!}\,
\sum_{n=0}^{\infty} \int_0^{\infty}\d t\,{\rm e}^{-n\,t}\,t^{2\,m+3},
\end{equation}
where, as we had done earlier, we have expressed ${\rm cos}(z\,t)$ and 
$(1-{\rm e}^{-t})^{-1}$ as a power series.
Evaluating the integral and carrying out the sum over $m$ leads to
\begin{equation}
\psi_3(i\,z)+\psi_3(-i\,z)
=6\,\sum_{n=0}^{\infty}\l[\f{1}{(z+i\,n)^4}+\f{1}{(z-i\,n)^4}\r].
\label{eq:psi3-f}
\end{equation}
Comparing Eqs.~(\ref{eq:psi3}) and (\ref{eq:psi3-f}) we arrive at the
following series representation of ${\rm cosech}^4(\pi z)$:
\begin{equation}
{\rm cosech}^4(\pi\,z)
=-\frac{2}{3\,\pi^2}\,
\l[\frac{1}{z^2}+\sum_{n=1}^{\infty}\frac{1}{(z+i\,n)^2}
+\sum_{n=1}^{\infty}\frac{1}{(z-i\,n)^2}\r]
+\frac{1}{\pi^4}\,\l[\frac{1}{z^4}
+\sum_{n=1}^{\infty}\frac{1}{(z+i\,n)^4}
+\sum_{n=1}^{\infty}\frac{1}{(z-i\,n)^4}\r],
\end{equation}
which is the result we have made use of in the text.


\section{Evaluating the integrals}\label{app:ei}

In this appendix, we shall outline the evaluation of the integrals 
$I_1(\alpha,t)$ and $I_2(\alpha,t)$ as described by Eqs.~(\ref{eq:I-at}).

\par

If we substitute $y'={\rm coth}\l[\pi\,(t'+i\epsilon)/\beta\r]$ in the expression 
for $I_1(\alpha,t)$, it reduces to
\begin{equation}
I_1(\alpha,t)
=-\frac{\beta}{\pi}\,
\int_{-1}^{y}\,\d y'\l(\frac{y'+1}{y'-1}\r)^{p-1}
\,\l[\f{1}{y'-1}+\f{1}{(y'-1)^2}\r],
\end{equation}
where $p=\alpha\,\beta/(2\,\pi)$.
If we now further set $u'=(y'+1)/(y'-1)$, we obtain that
\begin{equation}
I_1(\alpha,t)
=-\frac{\beta}{\pi}\, {\rm e}^{\alpha\, t}\,\int_{0}^{1}\d x\, x^{p-1}\,
\l[\frac{1}{1-{\rm e}^{2\,\pi\, t/\beta}\,(1+i\,\epsilon)\,x}-\frac{1}{2}\r].
\end{equation}
We can make use of the following integral representation of the hypergeometric 
function to evaluate the above integral~\cite{lebedev-1965}
\begin{equation}
{}_2F_1[a,b;c;z]
=\frac{\Gamma (c)}{\Gamma (b) \Gamma(c-b)}\,
\int_{0}^{1}\d x\, x^{b-1}\,(1-x)^{c-b-1}\,(1-z\,x)^{-a},
\label{eq:hgf-d}
\end{equation}
for ${\rm Re.}\,c>{\rm Re.}\,b>0$ and $|{\rm arg}(1-z)|<\pi$.
We find that $I_1(\alpha,t)$ can be written as
\begin{equation}
I_1(\alpha,t)
=\frac{{\rm e}^{\alpha\,t}}{\alpha}\,
\l\{1-2\,{}_2F_1\l[1,p;p+1;{\rm e}^{2\,\pi\,t/\beta}\,(1+i\,\epsilon)\r]\r\}.
\end{equation}
Similarly, we can evaluate $I_2(\alpha,t)$ to arrive at 
\begin{equation}
I_2(\alpha,t)
=-\frac{{\rm e}^{-\alpha\,t}}{\alpha}\,
\l\{1-2\,{}_2F_1\l[1,p;p+1;{\rm e}^{-2\,\pi\,t/\beta}\,(1+i\,\epsilon)\r]\r\}.
\end{equation}
Since the mean-squared displacement $\sigma^2(t)$ must be a real quantity, we 
write the integral $I_1(\alpha,t)$ as follows:
\begin{equation}
I_1(\alpha,t)
=\frac{{\rm e}^{\alpha\, t}}{\alpha}\,
\l\{1-{}_2F_1\l[1,p;p+1;{\rm e}^{2\,\pi\, t/\beta}\,(1+i\,\epsilon)\r]
-{}_2F_1\l[1,p;p+1;{\rm e}^{2\,\pi\,t/\beta}\,(1-i\,\epsilon)\r]\r\}.
\end{equation}
Upon writing the quantity $I_2(\alpha,t)$ in a similar fashion and 
substituting the resultant expressions in Eq.~(\ref{eq:msd-pv}),
we obtain that
\begin{eqnarray}
\sigma_{z}^2(t)
=\frac{12}{\alpha_1\,\alpha_2}\,
\biggl\{\gamma_{\rm E}+{\rm ln}\l[2\,{\rm sinh}(\pi\,t/\beta)\r]\biggr\}
+\frac{12}{\alpha_1\,(\alpha_1+\alpha_2)}\,F(p_1,t)
+\frac{12}{\alpha_2\,(\alpha_1+\alpha_2)}\,F(p_2,t),\label{eq:msd-a}
\end{eqnarray}
where the function $F(p,t)$ is defined as
\begin{eqnarray}
F(p,t)
&\equiv&\frac{1}{4\,p}\,
\l\{{}_2F_1\l[1,p;p+1;{\rm e}^{2\,\pi\, t/\beta}\,(1+i\,\epsilon)\r]
+{}_2F_1\l[1,p;p+1;{\rm e}^{2\,\pi\, t/\beta}\,(1-i\,\epsilon)\r]\r\}\nn\\
& &+\,\frac{1}{2\,p}\,{}_2F_1\l[1,p;p+1;{\rm e}^{-2\,\pi\,t/\beta}\r]
+\psi_0(p),\label{eq:F}
\end{eqnarray}
with $\psi_n(z)$ being the polygamma function~\cite{gradshteyn-2007}.
In order to write $F(p,t)$ more compactly we make use of the 
identity~\cite{abramowitz-1972}
\begin{eqnarray}
{}_2F_1\l[a,b;c;z\r]
&=&\frac{\Gamma(c)\,\Gamma(b-a)}{\Gamma(b)\Gamma(c-a)}\,
(-z)^{-a}\,{}_2F_1\l[a,1-c+a;1-b+a;z^{-1}\r]\nn\\
& &+\,\frac{\Gamma(c)\,\Gamma(a-b)}{\Gamma(a)\Gamma(c-b)}\,
(-z)^{-b}\,{}_2F_1\l[b,1-c+b;1-a+b;z^{-1}\r],
\end{eqnarray}
where $(a,b,c)\not\in\mathbb{Z}$ or $(a-b)\not\in\mathbb{Z}$ and 
$\vert{\rm arg}(-z)\vert<\pi$.
We find that $F(p,t)$ can be written as
\begin{equation}
F(p,t)
=\frac{\pi}{2}\, {\rm cot}(\pi p)\; {\rm e}^{-2\,\pi\,p\,t/\beta} 
+\frac{{\rm e}^{-2\,\pi\, t/\beta}}{2\,(1-p)}\,
{}_2F_1\l[1,1-p;2-p;{\rm e}^{-2\,\pi\,t/\beta}\r] 
+\frac{1}{2\,p}\, {}_2F_1\l[1,p;p+1;{\rm e}^{-2\,\pi\, t/\beta}\r]+\psi_0(p),
\end{equation}
which is the result we have made use of in the text.
We should clarify that since $p_1$ and $p_2$ are, in general, not integers 
[cf. Eqs.~(\ref{eq:p1-p2})], Eq.~(\ref{eq:msd-a}) is valid for all finite
values of the mass $m$ of the mirror and the inverse temperature $\beta$.
We have quoted the result~(\ref{eq:msd-a}) with $F(p,t)$ given by 
Eq.~(\ref{eq:F}) in the text.


\section{Divergence in the mean-squared displacement in the limit of
zero temperature}\label{app:div}

In this section, we shall discuss a subtle point concerning the zero 
temperature limit of the finite temperature result~(\ref{eq:msd-a}) 
for the mean-squared displacement of the mirror.

\par

We find that a logarithmic divergence arises if we blindly take the zero 
temperature limit (\ie\/ $\beta\to\infty$) of the final result~(\ref{eq:msd-a}) 
for the mean-squared displacement of the mirror at a finite temperature.
In the previous appendix, we had expressed the integrals integrals 
$I_1(\alpha,t)$ and $I_2(\alpha,t)$ in terms of the hypergeometric 
function using the definition~(\ref{eq:hgf-d}). 
Note that the representation~(\ref{eq:hgf-d}) is valid only for ${\rm Re.}\,c
>{\rm Re.}\,b>0$ and $|{\rm arg}(1-z)|<\pi$. 
Hence, for the expression~(\ref{eq:F}) describing $F(p,t)$ in terms of the 
hypergeometric functions to be valid, $p_1$ and $p_2$ should be positive 
definite for all values of $\beta$ and $m$. 
One can easily show that, while $p_1$ remains positive, $p_2$ tends to zero
in the limit of $\beta \to \infty$. 
Since [cf.~Eq.~(\ref{eq:hgf-d})]
\begin{equation}
\f{1}{p_2}{}_2F_1\l[1,p_2;p_2+1;z\r]=\int_0^1\,\d x\, x^{p_2-1}(1-z\,x)^{-1},
\end{equation}
we can write
\begin{equation}
\f{1}{p_2}{}_2F_1\l[1,p_2;p_2+1;z\r]
=\f{z}{p_2+1}{}_2F_1\l[1,p_2+1;p_2+2;z\r]+{\cal I}(p_2),
\label{eq:hgf-r2}
\end{equation}
where 
\begin{equation}
{\cal I}(p_2)
=\l\{\begin{array}{ll}
1/p_2 &{\rm when}\quad p_2>0,\\
-\ln\, \varepsilon &{\rm when}\quad p_2=0,   
\end{array}\r.
\end{equation}
with $\varepsilon \to 0$. 
On substituting Eq.~(\ref{eq:hgf-r2}) in Eq.~(\ref{eq:msd-a}) and making use 
of the following identity~\cite{abramowitz-1972}:
\begin{equation}
\psi_m(z)=\psi_m(z+1)+\f{(-1)^{m+1}\,m!}{z^{m+1}},
\end{equation}
we obtain that 
\begin{eqnarray}
\sigma_{z}^2(t)
=\frac{12}{\alpha_1\,\alpha_2}\,
\biggl\{\gamma_{_{\rm E}}+{\rm ln}\l[2\,{\rm sinh}(\pi\, t/\beta)\r]\biggr\}
+\frac{12}{\alpha_1\,(\alpha_1+\alpha_2)}\,F(p_1,t)
+\frac{12}{\alpha_2\,(\alpha_1+\alpha_2)}\,\l[F(p_2+1,t)+{\cal J}(p_2)\r],
\end{eqnarray}
where ${\cal J}(p_2)$ is given by
\begin{equation}
{\cal J}(p_2)
=\l\{\begin{array}{ll}
0 &{\rm when}\quad \beta>0,\\
-\ln\, \varepsilon &{\rm when}\quad \beta\to\infty.    
\end{array}\r.
\end{equation}
In other words, the expression~(\ref{eq:msd-a}) for the mean-squared displacement 
of the mirror at a finite temperature would diverge logarithmically if we naively 
consider the zero temperature limit.
\end{widetext}



\begin{thebibliography}{99}
\bibitem{reif-1965}
F.~Reif, {\sl Fundamentals of Statistical and Thermal Physics} 
(McGraw Hill, New York, 1965),\/ Secs.~15.5--15.10.
\bibitem{callen-1951}
H.~B.~Callen and T.~A.~Welton, Phys.\ Rev.\ {\bf 83}, 34 (1951).
\bibitem{kubo-1957}
R.~Kubo, J.\ Phys.\ Soc.\ Jap.\ {\bf 12}, 570 (1957).
\bibitem{kubo-1966}
R.~Kubo, Rep.\ Prog.\ Phys.\ {\bf 29}, 255 (1966).
\bibitem{sinha-1992}
S.~Sinha and R.~D.~Sorkin, Phys.\ Rev.\ B {\bf 45}, 8123 (1992).
\bibitem{jaekel-1993a}
M.~T.~Jaekel and S.~Reynaud, J.\ Phys.\ I (France), {\bf 3}, 339 (1993).
\bibitem{jaekel-1993b}
M.~T.~Jaekel and S.~Reynaud, Phys.\ Lett.\ A\ {\bf 172}, 319 (1993).
\bibitem{jaekel-1992}
M.~T.~Jaekel and S.~Reynaud, Quant.\ Opt.\ {\bf 4}, 39 (1992).
\bibitem{gour-1999}
G.~Gour and L.~Sriramkumar, Found.\ Phys.\ {\bf 29}, 1917 (1999).
\bibitem{alves-2003}
D.~T.~Alves, C.~Farina and P.~A.~M.~Neto, J.\ Phys.\ A\ {\bf 36}, 
11333 (2003).
\bibitem{wu-2005}
C.~H.~Wu and D.~S.~Lee, Phys.\ Rev.\ D\ {\bf 71}, 125005 (2005).
\bibitem{alves-2008}
D.~T.~Alves, E.~R.~Granhen, M.~G.~Lima, Phys.\ Rev.\ D.\ {\bf 77}, 125001 
(2008).
\bibitem{wang-2014}
Q.~Wang and W.~G.~Unruh, Phys.\ Rev.\ D\ {\bf 89}, 085009 (2014).
\bibitem{wang-2015}
Q.~Wang and W.~G.~Unruh, Phys.\ Rev.\ D\ {\bf 92}, 063520 (2015).
\bibitem{jaekel-1997}
M.~T.~Jaekel and S.~Reynaud, Rep.\ Prog.\ Phys.\ {\bf 60}, 863 (1997).
\bibitem{dodonov-2001}
V.~V.~Dodonov, Adv.\ Chem.\ Phys. {\bf 119}, 309 (2001).
\bibitem{moore-1970}
G.~T.~Moore, J.\ Math.\ Phys.\ (N.Y.) {\bf 11}, 2679 (1970).
\bibitem{dewitt-1975}
B.~S.~DeWitt, Phys.\ Rep.\ {\bf 19}, 295 (1975).
\bibitem{fulling-1976}
S.~A.~Fulling and P.~C.~W.~Davies, Proc.\ R.\ Soc.\ A\ {\bf 348}, 
393 (1976).
\bibitem{ford-1982}
L.~H.~Ford and A.~Vilenkin, Phys.\ Rev.\ D {\bf 25}, 2569 (1982).
\bibitem{itzykson-1980}
C.~Itzykson and J.-B.~Zuber, {\sl Quantum Field Theory} (McGraw Hill, 
New York, 1980).
\bibitem{birrell-1982}
N.~D.~Birrell and P.~C.~W.~Davies, {\sl Quantum Field Theory in Curved
Space}\/ (Cambridge University Press, Cambridge, England, 1982). 
\bibitem{phillips-2001-03}
N.~G.~Phillips, B.~L.~Hu, Phys.\ Rev.\ D\ {\bf 63}, 104001 (2001);
Phys.\ Rev.\ D\ {\bf 67}, 104002 (2003).
\bibitem{cho-2015}
T.~Cho and B.~L.~Hu, Class.\ Quantum Grav.\ {\bf 32}, 055006 (2015).
\bibitem{gradshteyn-2007}
I.~S.~Gradshteyn and I.~M.~Ryzhik, {\sl Table of Integrals, Series and
Products}\/ 7th ed. (Academic Press, San Diego, 2007).
\bibitem{lebedev-1965}
N.~N.~Lebedev, {\sl Special Functions and Their Applications}\/
(Prentice-Hall, New Jersey, 1965).
\bibitem{abramowitz-1972}
M.~Abramowitz and I.~A.~Stegun, {\sl Handbook of Mathematical Functions}\/
(Dover, New York, 1972).
\end{thebibliography}
\end{document}